\documentclass[10pt,twosides]{classe-cmb}
\usepackage{amssymb,amsbsy,amsmath,amsfonts,amssymb,amscd}
\usepackage{latexsym,euscript,exscale,epic,eepic,epsfig}
\usepackage[francais,english]{babel}
\usepackage{times}

\newcommand{\ba}{\begin{eqnarray}}
\newcommand{\ea}{\end{eqnarray}}
\newcommand{\be}{\begin{equation}}
\newcommand{\ee}{\end{equation}}
\newcommand{\bea}{\begin{eqnarray}}
\newcommand{\eea}{\end{eqnarray}}
\newcommand{\beq}{\begin{equation}}
\newcommand{\eeq}{\end{equation}}
\newcommand{\beqar}{\begin{eqnarray}}
\newcommand{\eeqar}{\end{eqnarray}}
\newcommand{\beqars}{\begin{eqnarray*}}
\newcommand{\eeqars}{\end{eqnarray*}}
\newcommand{\bc}{\begin{center}}
\newcommand{\ec}{\end{center}}
\newcommand{\ben}{\begin{enumerate}}
\newcommand{\een}{\end{enumerate}}
\newcommand{\bit}{\begin{itemize}}
\newcommand{\eit}{\end{itemize}}
\newcommand{\bw}{\begin{widetext}}
\newcommand{\ew}{\end{widetext}}
\newcommand{\bcl}{\begin{columns}}
\newcommand{\ecl}{\end{columns}}
\newcommand{\ex}{\mbox{e}}
\newcommand{\dd}{\mbox{d}}

\newcommand{\eg}{\emph{e.g.~}}

\newcommand{\apriori}{\emph{a priori~}}

\newcommand{\Ka}{{\cal K}}

\newcommand{\GN}{G_{_\mathrm{N}}}

\newcommand{\mrm}[1]{\mathrm{#1}}
\newcommand{\mcl}[1]{\mathcal{#1}}

\newcommand{\lb}{\left(}
\newcommand{\rb}{\right)}
\newcommand{\lsb}{\left[}
\newcommand{\rsb}{\right]}

\newcommand{\rcb}{\right\}}
\newcommand{\nn}{\nonumber}

\newcommand{\bm}{\mathbf}
\newcommand{\Mp}{\mathrm{M_{Pl}}}

\pdfoutput=1
\usepackage{bm}
\usepackage{hyperref}
\usepackage{color}
\usepackage[utf8]{inputenc} 
\hypersetup{,
    bookmarks=true,                                 
    unicode=false,                                     
    pdftoolbar=true,                                  
    pdfmenubar=true,                         	
    pdffitwindow=false,                              
    pdfstartview={FitH},                    	
    pdftitle={My title},                     		
    pdfauthor={Author},                          
    pdfsubject={Subject},                   
    pdfcreator={Creator},                   
    pdfproducer={Producer},              
    pdfkeywords={keyword1} {key2} {key3},    
    pdfnewwindow=true,                      
    colorlinks=true,                         
    linkcolor=red,                          
    citecolor=darkblue,                 
    filecolor=magenta,                  
    urlcolor=darkblue,                   
    linktocpage=true
}
\definecolor{blue}{rgb}{0.19,0.64,0.54}
\definecolor{reddish}{rgb}{0.65, 0.2, 0.2}
\definecolor{red}{rgb}{0.7,0.3,0.3}
\definecolor{darkgreen}{rgb}{0.2,0.7,0.3}
\definecolor{darkblue}{rgb}{0.3,0.40,0.48}
\definecolor{gray}{rgb}{.8,.8,.8}

\def\spose#1{\hbox to 0pt{#1\hss}}

\def\lta{\mathrel{\spose{\lower 3pt\hbox{$\mathchar"218$}}
     \raise 2.0pt\hbox{$\mathchar"13C$}}}
\def\gta{\mathrel{\spose{\lower 3pt\hbox{$\mathchar"218$}}
     \raise 2.0pt\hbox{$\mathchar"13E$}}}

\ComParit{S1296-2147}
\PIT{FLA}
\PXHY{????}
\Add{?}

\Volume{0} \Year{2015} \FirstPage{1} \LastPage{??}
\AuteurCourant{Lilley and Peter}
\TitreCourant{Bouncing alternatives}
\Journal
\Rubrique{Rubrique}{Heading}
\SousRubrique{Sous-rubrique}{Sub-Heading}
\PresentePar{}{} \Recu{}{}{}

\begin{document}

\TitleOfDossier{Inflation: theoretical and observational status}

\title{Bouncing alternatives to inflation\\
\vskip0.2cm
Rebond primordial comme alternative à l'inflation}

\author{Marc Lilley and Patrick Peter}

\address{Sorbonne Universités, UPMC Univ. Paris 06, UMR 7095, Institut Lagrange de
Paris, F-75014, Paris, France\\
CNRS, UMR 7095, Institut d'Astrophysique de Paris (${\cal
  G}\mathbb{R}\varepsilon\mathbb{C}{\cal O}$), F-75014, Paris, France}

\maketitle
\thispagestyle{empty}

\begin{Abstract}
{Although the inflationary paradigm is the most widely accepted
explanation for the current cosmological observations, it does not
necessarily correspond to what actually happened in the early
stages of our Universe. To decide on this issue, two paths can be
followed: first, all the possible predictions it makes must be derived
thoroughly and compared with available data, and second, all the
imaginable alternatives must be ruled out. Leaving the first task to
all other contributors of this volume, we concentrate here on the
second option, focusing on the bouncing alternatives and their
consequences. 

\hskip2cm
Quoique le paradigme inflationaire soit maintenant communément
accepté comme représen\-tant la meilleure explication des données
cosmologiques, il n'est pas pour autant possible de dire qu'une telle
phase soit avérée. Pour s'approcher d'une telle conclusion, on peut
suivre deux chemins différents~: on peut explorer les conséquences
de l'inflation pour la pousser dans ses derniers retranchements, ou
bien, au contraire, étudier en détail les alternatives possibles. La première
option faisant l'objet de la plupart des contributions de ce volume,
nous nous concentrons ici sur la seconde, et présentons les modèles
dans lesquels une phase de contraction est suivie d'un rebond
conduisant à notre époque d'expansion.}
\end{Abstract}

\par\medskip\centerline{\rule{2cm}{0.2mm}}\medskip
\twocolumn
\setcounter{section}{0}

\section{Introduction}
Starting out in a dense state some 13.8 billion years ago, our
Universe and its evolution since this initial time are well understood, with an initially almost scale-invariant, but not
quite, spectrum of primordial perturbations condensing into the
presently observed large scale structures by means of gravitational
collapse. The very high densities of the early stages provide initial
conditions to explain the relative amounts of different nuclei, and
the ensuing phases, being controlled by well-known physical
mechanisms, permit to reconstruct, from the cosmic microwave
background (CMB) observations, the properties of the last scattering
surface. We have arrived at the point
\cite{Mukhanov:2005sc,PeterUzan2009} where cosmological data can be 
used to probe the earliest conceivable phases.

The most widely accepted paradigm for describing the earliest phases
of the Universe, when the energy density was a mere few orders of
magnitude below the Planck scale, is inflation
\cite{Lemoine:2008zz,Martin:2014vha}. Easily
implemented by means of a scalar field, this almost exponentially
expanding era rapidly leads to a flat Friedmann Lema\^\i{}tre (FL)
spacetime with a very slightly reddish spectrum of initial
perturbations from which the rest of the history of the Universe
ensues. As is, such a scenario is compatible with all currently
available data.

This contribution reviews some properties of some non inflationary
bouncing models. The first natural question that comes to mind before
going any further is: why should we bother with possible alternatives
to a working scenario? There are in fact many reasons, the first of
which being that the phase of inflation is silent relative to the
primordial singularity, as we discuss in Sec.~\ref{sec:sing}
below. The second is that there is no way we will ever be able to
assert that a phase of inflation did actually take place except
through its presently observable consequences. But then the question
arises as to whether other competing theories could induce similar
consequences. Thuus, examining all plausible scenarios in detail seems
to be the only way to assert whether inflation is the unique
possibility leading to our observable Universe. In the end, ruling out
 alternatives, or not, increases or decreases our level of confidence
 in inflation until it becomes, if ever, recognized as valid beyond any reasonable doubt. As
we shall see in Sec.~\ref{sec:puz}, there are bouncing alternative
explanations to the
standard model puzzles of homogeneity, flatness, isotropy, horizon and
the overproduction of relics, as well as many models, some of which
are listed in Sec.~\ref{sec:models}, in which those bounces can be
implemented.

Getting a background-compatible model is however not the end of the
story: the recently released P{\footnotesize LANCK} data
\cite{Planck:2013jfk,Ade:2013zuv} confirm what was suggested by previous
experiments, namely that the spectrum of primordial perturbations was
almost scale invariant: slightly red, with a spectral index
$n_\mathrm{s} =0.9639\pm 0.0047$, excluding exact scale invariance at
the 5$\sigma$ level. The level of non-gaussianity is compatible with
zero, and the contribution of tensor modes remains below
the $\sim\!10\%$ limit relative to the scalar amplitude. All these facts are
compatible with the perturbations having been produced by quantum
vacuum fluctuations of a single scalar degree of freedom, a natural
consequence of slow-roll single-field inflation.  Can a
non-inflationary bouncing reproduce such results? As of now, there is no
definite answer to this question. For this reason, and for lack of
space in the present article, we shall not discuss these points
below, and instead refer the reader to a recent review \cite{Battefeld:2014uga}
in which all the relevant constraints for the models exhibited
below are derived.

\section{The singularity} \label{sec:sing}

The fact that cosmology, or at least its classical implementation in
terms of general relativity (GR), always leads to the existence of
singularities stems from the well-known singularity theorems
\cite{Hawking:1973uf}. A general argument was proposed in
Ref.~\cite{Borde:2001nh}: in an FL spacetime with metric
\begin{eqnarray}
\dd s^2 =& \hskip-1cm -\dd t^2 + a^2(t)
\gamma^\Ka_{ij}\left(\bm{x}\right) \dd x^i \dd x^j
\nonumber\\ =& \hskip-2mm a^2(\eta) \left[ -\dd \eta^2 +
  \gamma^\Ka_{ij} \left(\bm{x}\right) \dd x^i \dd x^j\right],
\label{FLmetric}
\end{eqnarray}
with $\gamma^\Ka_{ij}$ the constant-curvature ($\Ka=0,\pm 1$) spatial
metric, let $\mathcal{U}^\mu \equiv \dd x^\mu/\dd \lambda$, with
$\lambda$ an affine parameter, be a lightlike tangent to a geodesic
curve, i.e. $\mathcal{U}^\mu\mathcal{U}_\mu = \gamma^\Ka_{ij} a^2
\mathcal{U}^i \mathcal{U}^j - \left( \mathcal{U}^0\right)^2=0$ and
$\mathcal{U}^\mu\nabla_\mu \mathcal{U}^\alpha=0$. Expanding the
geodesic equation in terms of the connections associated with the
metric (\ref{FLmetric}) and taking into account the lightlike
character of $\mathcal{U}$, one finds that
$$
\frac{\dd \mathcal{U}^0}{\dd\lambda} + H \left( \mathcal{U}^0\right)^2
= 0,
$$
which implies that
\begin{equation}
\frac{\dd}{\dd\lambda} \left( \frac{\dd t}{\dd \lambda} \right)
+H \left( \frac{\dd t}{\dd\lambda}
\right)^2 =0,
\label{geodesic}
\end{equation}
where the Hubble scale is $H=\dd \ln a/\dd t$. Eq.~(\ref{geodesic}) is
solved by choosing the affine parameter $\lambda$ to satisfy $\dd
\lambda = [a(t)/a(t_\mathrm{f})] \dd t$, with $t_\mathrm{f}$ a
reference time, today say. We now assume our spacetime to begin at
some initial coordinate time $t_\mathrm{i}$, which can take any value between $0$ say, to $-\infty$; this depends on the actual cosmological
realization. The average Hubble rate along the geodesic parameterized
by $\lambda$ is found to be
\begin{eqnarray}
H_\mathrm{average} \equiv& \hskip-3mm \displaystyle
\frac{1}{\lambda(t_\mathrm{f})-\lambda(t_\mathrm{i})}
\int_{\lambda(t_\mathrm{i})}^{\lambda(t_\mathrm{f})} H(\lambda)
\dd\lambda\cr =& \hskip-3mm \displaystyle
\frac{1}{\lambda(t_\mathrm{f})- \lambda(t_\mathrm{i})} \left\{
1-\frac{a\left[\lambda(t_\mathrm{i})\right]}{a\left[\lambda(t_\mathrm{f})\right]}
\right\}\cr \leq & \hskip-26mm \displaystyle
\frac{1}{\lambda(t_\mathrm{f})-\lambda(t_\mathrm{i})},
\label{Hav}
\end{eqnarray}
so that in order for $H_\mathrm{average}$ to be strictly positive, a
condition which is generally satisfied in inflationary models, one
finds that the interval in affine parameter must be finite, and
therefore that the spacetime under consideration is not geodesically
complete. This argument cane be extended to timelike geodesics and more
arbitrary cosmological models, i.e. with no specific assumptions
regarding homogeneity and isotropy. This requires the definition of a local
expansion rate that is not dependent on the special FL metric solution; in
this case, it is the deviation between neighboring geodesics that
needs be used explicitly to define the expansion rate (in the highly
symmetric FL universe, the geodesic deviation is given by the
expansion only, as this is the only relevant observable). The
conclusion then is that, regardless of any energy condition, inflating
spacetimes are past incomplete.

An obvious way out of this problem consists in allowing the average
Hubble rate to be negative. This amounts to having some amount of
contraction, and hence, given that we observe the Universe to be currently
expanding, that it must have bounced. In the framework of GR however,
this is not always easy.  

\begin{figure}[h!]
\centering
\includegraphics[height=7.3cm]{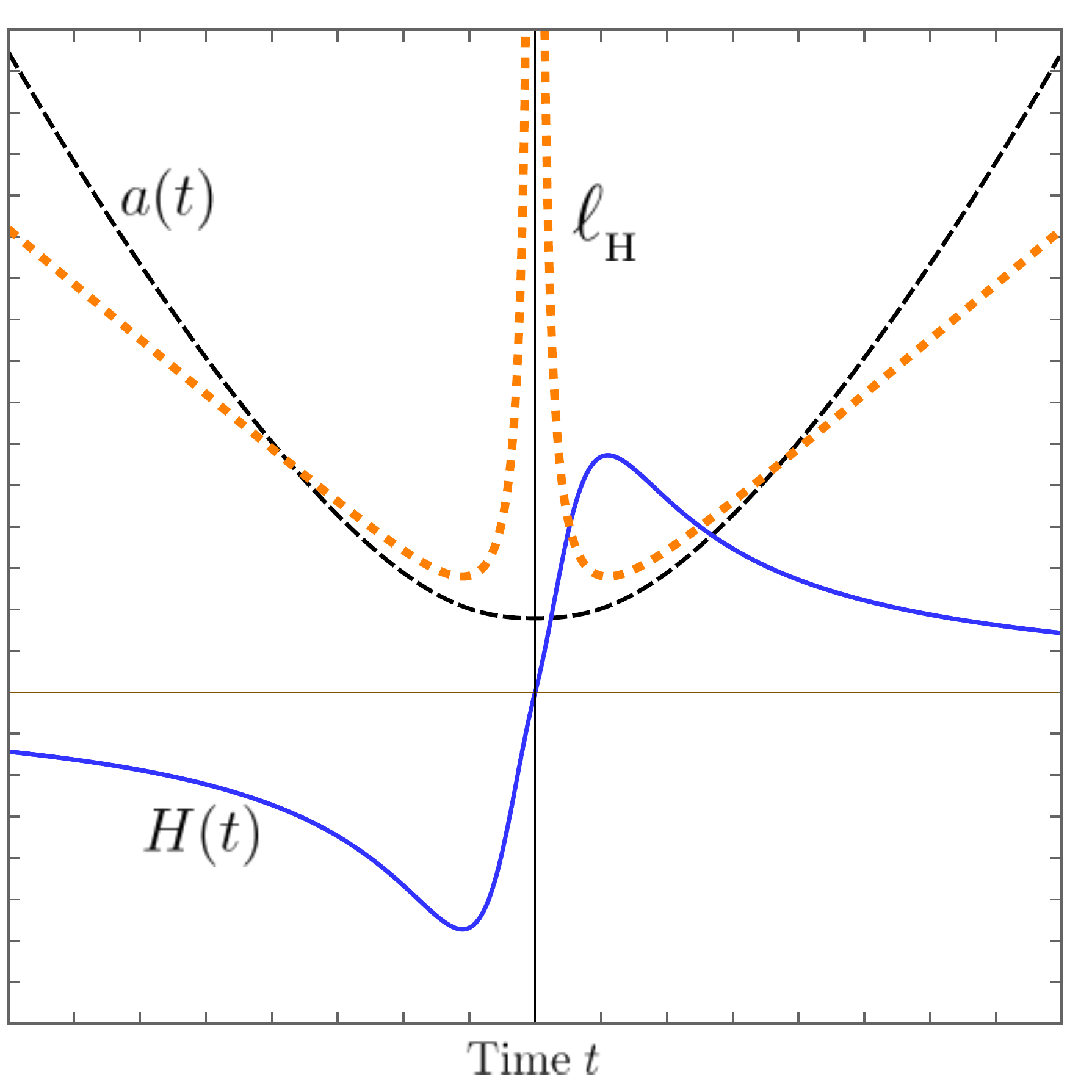}
\caption{ \label{aH} \footnotesize Typical time evolution of the scale
  factor $a(t)$ (dashed line), Hubble rate $H(t)\equiv \dot a /a$
  (full line) and Hubble length $\ell_{_\mathrm{H}}(t)=H^{-1}$ (dotted
  line) for a bouncing scenario. For large negative times
  (conventionally setting the bounce at $t=0$), the scale factor
  decreases in a non-accelerated way, then it curves up, accelerating
  and rendering the curve convex, finally connecting, not necessarily
  in a symmetric way, to a more standard non-accelerating
  expansion. The Hubble rate starts vanishingly small, then decreases
  to large negative values, passes through the bounce almost linearly
  increasing, reaches a maximum and then decreases back to its usual
  behavior. The Hubble length is originally very large, reaches a
  minimum and diverges at the bounce point: there is no super-Hubble
  scale at the bounce!}
\end{figure}

Using the metric (\ref{FLmetric}) and a fluid stress-energy tensor
$T_{\mu\nu} = \left(\rho + P\right) u_\mu u_\nu + P g_{\mu\nu}$ with
energy density $\rho$, pressure $P$, and $u_\mu$ a timelike vector,
the Einstein equations read
\begin{equation}
H^2+\frac{\Ka}{a^2} = \frac13\rho, \ \ \ \ \ \dot H+H^2 = \frac{\ddot
  a}{a} =-\frac16\left(\rho + 3 P\right),
\label{FriedCosm}
\end{equation}
leading to
\begin{equation}
\dot H = \frac{\Ka}{a^2} -\frac12 \left( \rho+P\right),
\label{Hdot}
\end{equation}
(we use natural units where $\hbar=c=8\pi \GN\equiv 1$ so that the
Planck mass $\Mp \equiv \GN^{-1/2}$ is dimensionless).  Although
having an inflationary phase with $\ddot a >0$ merely demands
the violation of the Strong Energy Condition (SEC: $\rho + 3P>0$), a
bounce, requiring $H\to 0$ while $\dot H>0$ at the same time, implies
that either the spatial sections must be positively curved ($\Ka>0$)
or the Null Energy Condition (NEC: $\rho+P>0$) must be violated. In
the former case, the scale
factor at the bounce $a_{_\mathrm{B}}$ is obtained as the
solution of $3 \Ka/a_{_\mathrm{B}}^2 = \rho(a_{_\mathrm{B}})$ and
must satisfy $\Ka/a_{_\mathrm{B}}^2 > - P(a_{_\mathrm{B}})$. This
condition is for instance fullfilled in the very simple case in which
a single scalar field evolves in a potential with a local maximum.
\cite{Martin:2003sf,Falciano:2008gt}.  The bouncing solution, seen
generically as indicated on Fig.~\ref{aH}, does have an accelerating
phase, the scale-factor curve being convex at the bounce; although
this technically implies the SEC to be also violated during a bouncing
epoch, this cannot be understood as an inflating phase since the
accelerating phase is not associated with a large increase of the
size of the Universe.

In the more familiar (to inflation-oriented cosmologists) case of
vanishing or negligible spatial curvature\footnote{Note that this is an
  assumption that can only be checked a posteriori: given a material
  content with positive and negative energy components, one must first
  solve the Friedmann equation for the minimum scale factor
  $a_{_\mathrm{B}}$, and then verify that the curvature term
  $\Ka/a_{_\mathrm{B}}^2$ is indeed negligible with respect to all
  other contributions.}, as mentioned above, Eqs.~(\ref{FriedCosm}) imply a much more
stringent constraint, namely that the NEC be violated; as discussed in Sec.~\ref{sec:models}, this
often leads to various instabilities which then need to be tamed in order
for the model to make any sense at all.

\section{Standard model puzzles, bouncing solutions -- new issues}
\label{sec:puz}

The reason why the inflationary scenario is so fashionable stems
from its successes in solving the standard hot big bang puzzles in a
unified way, while at the same time providing a means of producing
perturbations whose spectrum can be made to agree with all known
data. Can a bouncing scenario, on top of naturally avoiding the
singularity, propose satisfying solutions to the standard hot big bang
puzzles?  If it is the case, can a bouncing scenario provide a means
to generate cosmological perturbations whose statistics agree with
observations?  As mentioned before, we refer the reader to
ref.~\cite{Battefeld:2014uga} for a detailed discussion of this latter
question, and focus in the remainder of this review on bouncing
solutions to the background cosmological problems and on a review of 
existing bouncing models.

\subsection{Horizon and flatness puzzles} \label{puzzles}

The standard Hot Big-bang model suffers from a few puzzling
problems, and we will treat in this section how a bounce, which
implies a contracting phase preceding the current expansion, 
deals with the two most important, namely the horizon and flatness
problems. 
We refer the reader to Refs.~\cite{Battefeld:2014uga,Peter:2008qz} for
more details relative to the other commonly addressed puzzles.

\paragraph*{$-$ {\it Horizon}\\\\}

The horizon problem relates to the inability to explain the
quasi-homogeneity of the observable universe within the context of
standard cosmology in which the entire evolution of the universe
consists in decelerated expansion during the radiation and matter dominated
epochs.  In the context of
inflation, the necessity of a period of accelerated expansion for a
period lasting a
minimum of $N\sim 60$ e-folds can be phenomenologically
understood by computing the
solid angle subtended by causally connected regions. Assuming an
expansion history in which the universe is initially
radiation-dominated, then dominated by the inflaton, represented
by a fluid $X$ with equation of
state
parameter $w_{\scriptscriptstyle X}$ until a redshift $z_{\mrm e}$, then once again
radiation-dominated until the last
scattering surface at $z_{\mrm{lss}}$ and finally matter dominated till
today, we find that~\cite{Martin:2003bt}
\begin{align}
\Delta \Omega = & \displaystyle\frac{1}{2}\lsb 1-\lb
1+z_{\mrm{lss}}\rb^{-1/2}\rsb^{-1}\lb 1+z_{\mrm{lss}}\rb^{-1/2} 
\bigg\{ 1 \cr
& \left.+\displaystyle\frac{1-3w_{\scriptscriptstyle X}}{1+3w_{\scriptscriptstyle X}}\frac{1+z_{\mrm{lss}}}{1+z_{\mrm{e}}}\lsb 1
-\ex^{-N(1+3w_{\scriptscriptstyle X})/2}\rsb\rcb ,
\end{align}
where $N=\ln (a_{\mrm{e}}/a_{\mrm{i}})$ with $a_{\mrm{i}}$ the scale
factor at the onset of the $X$-dominated period.
If we assume that $N=0$, we recover standard cosmology and
$\Delta\Omega \sim 0.85$ degrees.  This corresponds to a total of
about $10^6$ causally disconnected regions in which, weirdly enough,
the CMB is everywhere the same up to 1 part in $10^5$. Increasing
$\Delta \Omega$ is easily achieved in the context of inflation by
requiring $w_{\scriptscriptstyle X}<-1/3$ and large positive $N$.

A similar calculation can be done in the context of bouncing
cosmology. Here again, we shall assume a phase dominated by a fluid
with equation of state parameter $w_{\scriptscriptstyle X}$ during
which the Universe first contracts.  The bounce is assumed
non-singular, occurring at a redshift $z_{\mrm b}$ and short enough that
we can ignore its contribution. It is followed by the standard radiation and
matter dominated phases. The solid angle subtended by
causally connected regions is then
\begin{align}
\Delta\Omega =  &\frac{1}{2}\lsb 1-\lb
1+z_{\mrm{lss}}\rb^{-1/2}\rsb^{-1}\lb 1+z_{\mrm{lss}}\rb^{-1/2}
\times\cr
&\bigg\{ 1 + \frac{1+z_{\mrm{lss}}}{1+z_{\mrm{b}}}\bigg[
\frac{3\left( 1+w_{\scriptscriptstyle X}\right)}{1+3w_{\scriptscriptstyle X}}
\ex^{-N(1+3w_{\scriptscriptstyle X})/2} \cr & - \frac{2\left(2+3w_{\scriptscriptstyle X}\right)}{1+3w_{\scriptscriptstyle X}}\bigg]+1
\bigg\}.
\end{align}
In this expression, $N=\ln(a_{\mrm{i}}/a_{_\mathrm{B}}) \leq 0$ with
$a_{\mrm i}$ and
$a_{_\mathrm{B}}$ the scale factors at the onset of the $X$-dominated
contraction and at the bounce respectively.  Here, large values of
$\Delta\Omega$ are obtained for $w>-1/3$ and large values of
$|N|$. Thus, in contrast with inflation, the horizon problem can be
solved using a fluid that satisfies all energy conditions.

\paragraph*{$-$ {\it Flatness}\\\\}

The flatness problem is easily understood by working with the
ratio of
the Friedmann equation with the critical density
$\rho_{\mrm{crit}}=3H^2$, which is the energy density
the Universe would have if it had exactly flat spatial sections.
In the presence of the spatial curvature term, the Friedmann
equation (\ref{FriedCosm}) takes the form
\begin{equation}
\frac{\Ka}{a^2H^2}=\sum_{i=1}^N\Omega_i-1=\Omega_{_\mathrm{T}}-1,
\label{eq:FriedCrit}
\end{equation}
with
\begin{equation}
\Omega_i=\frac{\rho_i}{\rho_{\mrm{crit}}}=\frac{H^2(t_0)}{H^2(t)}\Omega_i(t_0)\lb
\frac{a}{a_0}\rb^{-3(1+w_i)}.
\end{equation}
From observations, we know that the total density parameter today is
$\Omega_{_\mathrm{T}}(t_0)\simeq 1$. It is easy to recast
Eq.~(\ref{eq:FriedCrit}) in the convenient form~\cite{Martin:2003bt}
\begin{align}
\Omega_{_\mathrm{T}}(a)=&\sum_{j=1}^N\Omega_j(t_0) \lb \frac{a}{a_0} \rb^{-3(1
+w_j)}
\times\cr
&\bigg\{\sum_{i=1}^N \Omega_i(t_0) \lb \frac{a}{a_0} \rb^{-3(1
+w_i)}-\cr &\left[ \Omega_{_\mathrm{T}}(t_0)-1\right] \lb \frac{a}{a_0}\right)^{-2}
\bigg\}^{-1}.
\label{Omegat}
\end{align}

At early times (small scale factor), the Universe is radiation dominated,
and Eq.~(\ref{Omegat}) simplifies to
\be
\Omega_{_\mathrm{T}}(t)-1\simeq
\frac{\Omega_{_\mathrm{T}}(t_0)-1}{\Omega_{\mrm{rad}}(t_0)}\lb
\frac{1}{1+z}
\rb^2.
\label{eq:eq1}
\ee
For $z \gg 1$,
$\Omega(t)-1$ must be much less than 1.  For instance, taking
$z_{\mrm{nucl}}=3\times 10^8$, $\Omega_{\mrm{rad}}(t_0)=10^{-4}$,
and $\Omega_{_\mathrm{T}}(t_0)-1=0.01$, one
finds $\Omega_{_\mathrm{T}}(t_{\mrm{nucl}})-1\sim 10^{-15}$.  The value of
the total density parameter at nucleosynthesis required to satisfy
today's observed value $\Omega_{_\mathrm{T}}(t_0)\sim 1$ is highly fine-tuned and
thus highly improbable. This embodies the flatness problem of standard
cosmology.\\

Let us now consider the case of a bouncing universe which contracts in
a phase dominated by a fluid of equation of state parameter
$w_{\scriptscriptstyle X}$,
bounces and then expands according to the standard scenario.
Note that the set of equations above do not apply at the bounce
point
where $H=0$. In fact, in the presence of a spatial curvature term,
$\Omega_{_\mathrm{T}}$
diverges at the bounce.

For a universe dominated by some fluid $X$, one has
\be
\Omega_{_\mathrm{T}}(t)=\frac{\Omega_{\scriptscriptstyle X}(t_\mathrm{i})}{\Omega_{\scriptscriptstyle X}(t_\mathrm{i})-\left[\Omega_{_\mathrm{T}}(t_\mathrm{i})-1
\right]\lb a/a_\mathrm{i}\rb^{1+3w_{\scriptscriptstyle X}}}.
\ee
The total density parameter at the end of the contracting phase at $t^-$
is given by 
\be
\Omega_{_\mathrm{T}}(t_-)-1\simeq \frac{\Omega_{_\mathrm{T}}(t_\mathrm{i})-1}{\Omega_X(t_\mathrm{i})}\lb
\frac{a^-}{a_\mathrm{i}}\rb^{1+3w_{\scriptscriptstyle X}},
\ee
while it is given by Eq.~(\ref{eq:eq1}) at the beginning of the
expanding phase, for $t=t_+$ and $z=z_+$.
The difference 
\be
\Delta\Omega_{_\mathrm{T}}=\lsb\Omega_{_\mathrm{T}}(t_+)-1\rsb-\lsb\Omega_{_\mathrm{T}}(t_-)-1\rsb
\ee
can be computed using Eq.~(\ref{eq:FriedCrit}) and the Taylor
expansion of the scale factor close to the bounce,
\be
a(t)=a_{_\mrm{B}}\lsb 1 + \lb\frac{t}{t_{\mrm{c}}}\rb^2 + \beta \lb
\frac{t}{t_{\mrm{c}}}\rb^3+\dots\rsb\,.
\ee
One finds
\be
\Delta\Omega_{_\mathrm{T}}\simeq -\frac{3\beta}{2} \lb
\frac{t_{\mrm{c}}}{a_{_\mrm{B}}}\rb^2.
\ee
Generically, in the absence of any fine-tuning, one should assume
$\Omega_{_\mathrm{T}}(t_\mathrm{i})-1$ and
$\Omega_X(a_\mathrm{i})$ take values of $\mathcal{O}(1)$ while it is known that
$\Omega_{\mrm{rad}}(t_0)\simeq 10^{-4}$ and $\Omega_{_\mathrm{T}}(t_0)-1\leq
10^{-2}$.  Hence we have
\be
\lb\frac{a_-}{a_\mathrm{i}}\rb^{1+3w_{\scriptscriptstyle X}}-\frac{3\beta}{2}
\lb\frac{t_{\mrm{c}}}{a_{_\mrm{B}}}\rb^2\leq 10^6\times z_+^{-2}\,.
\ee
Taking as before $z_+\simeq 10^{28}$, and with $w=1/3$, we have
\be
\ex^{2N}-\frac{3\beta}{2}
\lb\frac{t_{\mrm{c}}}{a_{_\mrm{B}}}\rb^2 \leq 10^{-50},
\ee
where $N<0$. Thus, for $N\leq -60$, and $\beta$ of order $1$,
a bounce with a short characteristic timescale and a large value of the
scale factor at the bounce such that $t_{\mrm{c}}/a_{_\mrm{B}} \leq
 10^{-25}$ can satisfy current constraints on the spatial curvature of
 the universe.
 
\subsection{Shear/BKL instability} \label{shear}

In a bouncing scenario, the standard puzzles find natural solutions
because of the contracting phase. However, such a phase can also
induce another problem: the fate of any initial
amount of anisotropy. To focus on this question, we consider a
spatially flat model whose dynamics derives from the Bianchi I metric
$\dd s^2 = -\dd t^2 + a^2(t)\dd\bm{x}^2$, whose spatial part reads
\begin{equation}
\dd\bm{x}^2 =\ex^{2\theta_x (t)} \dd x^2 + \ex^{2\theta_y (t)} \dd y^2
+ \ex^{2\theta_z (t)} \dd z^2,
\label{BianchiI}
\end{equation}
with $\sum_i \theta_i \equiv \theta_x+\theta_y+\theta_z=0$.  Plugging
(\ref{BianchiI}) into the Einstein equations generalizes (\ref{FriedCosm}) to
\begin{equation}
H^2 \equiv \left(\frac{\dot{a}}{a}\right)^2= \frac13\rho +\frac16
\sum_i\dot\theta^2_i \equiv \frac13\left( \rho + \rho_\theta\right)
\label{H2theta}
\end{equation}
and
\begin{equation}
\dot H = -\frac12 \left( \rho + P\right) -\frac12
\sum_i\dot\theta^2_i,
\label{dotHtheta}
\end{equation}
where we have identified the shear energy density $\rho_\theta$
contained in the anisotropy stemming from the functions $\theta_i$:
Eqs.~(\ref{H2theta}) and (\ref{dotHtheta}) imply that $\ddot \theta_i
+ 3 H \dot \theta_i=0$, and therefore $\rho_\theta\propto a^{-6}$.

With dust and radiation scaling as $\rho_\mathrm{m} \propto a^{-3}$
and $\rho_\mathrm{r} \propto a^{-4}$ respectively, the above result is
a catastrophe: as the universe contracts, any initial anisotropy,
however small\footnote{Actually, the problem only arises in the
  presence of primordial classical shear: it has been shown that if
  the primordial shear is generated by quantum vacuum fluctuations,
  scalar and vector perturbations remain comparable
  \cite{Vitenti:2011yc,Pinto-Neto:2013zya}.}, will grow until it
eventually dominates the dynamics.  This was shown
\cite{Belinsky:1970ew} by Belinsky, Khalatnikov and Lifshitz (BKL) to
induce an instability sufficient to spoil the bounce.

One way out of the shear problem is to add an extra component, usually
a scalar field in a potential satisfying specific constraints, with
large effective equation of state $w_\phi\gg 1$, so that the resulting
Friedmann equation reads
\begin{equation}
H^2=\frac{1}{3}\left[-\frac{3\Ka}{a^2}+\frac{\rho_{\mathrm{m}0}}{a^3}+
  \frac{\rho_{\mathrm{r}0}}{a^4}+ \frac{\rho_{\theta 0}}{a^6} 
  +\frac{\rho_{\phi0}}{a^{3(1+w_{\phi})}}\right].
\end{equation}
If this so-called ekpyrotic phase~\cite{Khoury:2001wf}, lasts long en-
ough, it
eventually comes to dominate over all other consti\-tuents when $a\to
0$, including the shear contribution. The Universe then bounces and
starts expanding again while in a fully symmetric FL phase, a
condition absolutely required to explain the observational data.

The bounce itself is another matter, which we now turn to.

\section{Existing models}
\label{sec:models}

There exists a large number of bouncing cosmological models
in the literature; we shall refer the reader to Ref.~\cite{Battefeld:2014uga}
for an exhaustive review and all the relevant references. We will
instead focus here on a few models and give concrete examples.

\subsection{Classical bounces}

Bouncing models predate by many decades the inflationary
paradigm, as they were first introduced in the 1930's. Classical
models involve unconventional perfect fluids or scalar fields with
possibly non standard kinetic terms, or various combinations of
those. The most conservative setup that may be used to obtain a
bounce is to introduce spatial curvature and to violate the
strong energy condition. In such a setup, scalar field matter is
required in order to achieve $\rho+3P<0$ and either a
quasi-symmetric bounce or a phase of inflation is needed to drive
$\Omega_{\mcl K}$ towards zero.

\paragraph*{$\bullet$ {\bf Perfect fluids}\\\\}

In a theory restricted to GR and FL spacetime, generically,
for $\Ka\ne +1$, the null energy condition has to be violated in
order to obtain a bounce, as discussed below Eq.~(\ref{Hdot}).
Exotic hydrodynamical fluids that violate the null energy
condition, and thus all other energy conditions, are \apriori
allowed, and models using those can be built in the
framework of an FL spacetime. Assuming an expansion
for the  scale factor of the form
\begin{equation}
a=a_0+b\eta^{2n}+d\eta^{2n+1}+e\eta^{2n+2},
\end{equation}
where $n\geq 1$, and $(a_0,b,d,e)$ constant, it is possible to compute,
in a fully analytic way,
the evolution of {\it adiabatic} perturbations around the
bounce \cite{PP01}.  The possible choices for $n$ and $\Ka$ are: (i)
$n>1$ and $\Ka \ne 0$; (ii) $n>1$ and $\Ka=0$; (iii) $n=1$ and
$d\ne 0$ $\forall\, \Ka$; (iv) $n=1$ and $d=0$ $\forall\, \Ka$;
setting $d=0$ restricts to a symmetric bounce.  In the
first three cases, the Bardeen potential $\Phi$ describing the
gauge-invariant perturbations turns out to be singular at the
bounce, while in case (iv), although $\Phi$ is well behaved, the NEC
needs be violated even if $\Ka=+1$.  Given that at late
times, it has to be satisfied, there must exist a time
$t_*$ at which $\rho(t_*)+P(t_*)=0$. It turns out that, at this
{\it NEC transition}, the growth of $\Phi$ is unlimited, raising
potential questions on the perturbative expansion through a bouncing
phase.

When entropy perturbations are considered in addition to the
adiabatic ones, they are found to be sourced by the interaction
of the hydrodynamical fluids involved in
the cosmological evolution. This implies that the fluids do not evolve
independently.  Inclusion of entropy perturbations has the
effect of regularizing the Bardeen potential and its derivatives
at the NEC transition. It
may thus be concluded that perfect fluid dominated
bouncing models in which both
adiabatic and entropy perturbations are taken into account are regular
and do not necessarily
lead to strong backreaction effects of the perturbations
onto the background geometry \cite{Peter:2003rg}.

\vskip5mm

When it comes to violating energy conditions, it is tempting to
make use of scalar fields: most implementations of the
inflationary paradigm are based on scalar fields, and the bounce
does no better in that respect! Bouncing models powered by
such fields can be broadly distinguished in two categories,
depending on the coupling (minimal or extended) with the
geometry. We now discuss both these possibilities.

\paragraph*{$\bullet$ {\bf Minimally coupled scalar fields theories}\\\\}

Theories using minimally coupled scalar fields can be
separated in two categories.  First, an ordinary scalar
field, with standard kinetic term and a potential. In this case, in
order to obtain a bouncing cosmology and preserve the weak energy
condition, one needs $\Ka=+1$. We shall not dwell here with
such cases, which then demand the curvature problem to be
addressed independently, and may result in perturbations
being large and potentially highly non Gaussian \cite{Gao:2014eaa}.

The second category naturally involves non standard kinetic terms;
those can be generalized to the Galileon theories (non minimal coupling,
see below). Their advantage over the standard kinetic terms is that
those can be implemented in a flat FL universe.

\paragraph*{$-$ {\it Ghost condensates}\\\\}

The simplest possible example of a non standard kinetic terms consists
in merely switching its sign, making it a so-called ghost field, which
yields an untenable theory because instabilities, both classical and
quantum, will immediately develop and ruin any configuration. Although
one can use a ghost as an effective means to initiate a bounce with
$\Ka=0$ \cite{PP02}, it makes more sense to induce NEC
violations with dynamical ghosts. This can
be achieved in
higher-derivative theories with second order equations of motion that
by construction prevent gradient or ghost instabilities. This is the
ghost condensate mechanism \cite{ArkaniHamed:2003uy}, whose
features we can sketch with the Lagrangian
\begin{equation}
\mcl L=P(X)\ \ \text{where}\ \  X\equiv
-\frac{1}{2}g^{\mu\nu}\partial_{\mu}\phi\partial_{\nu}\phi,
\label{PX}
\end{equation}
and where the pressure $P$ is an arbitrary function of the kinetic
energy $X$. Eq.~(\ref{PX}) in a flat FL metric yields
\begin{equation}
\frac{\dd}{\dd t}(a^3P_{,X}\dot\phi)=0,
\label{scalareom}
\end{equation}
where $P_{,X}\equiv\dd P/\dd X$.

\begin{figure}[h!]
\begin{center}
\includegraphics[width=7cm]{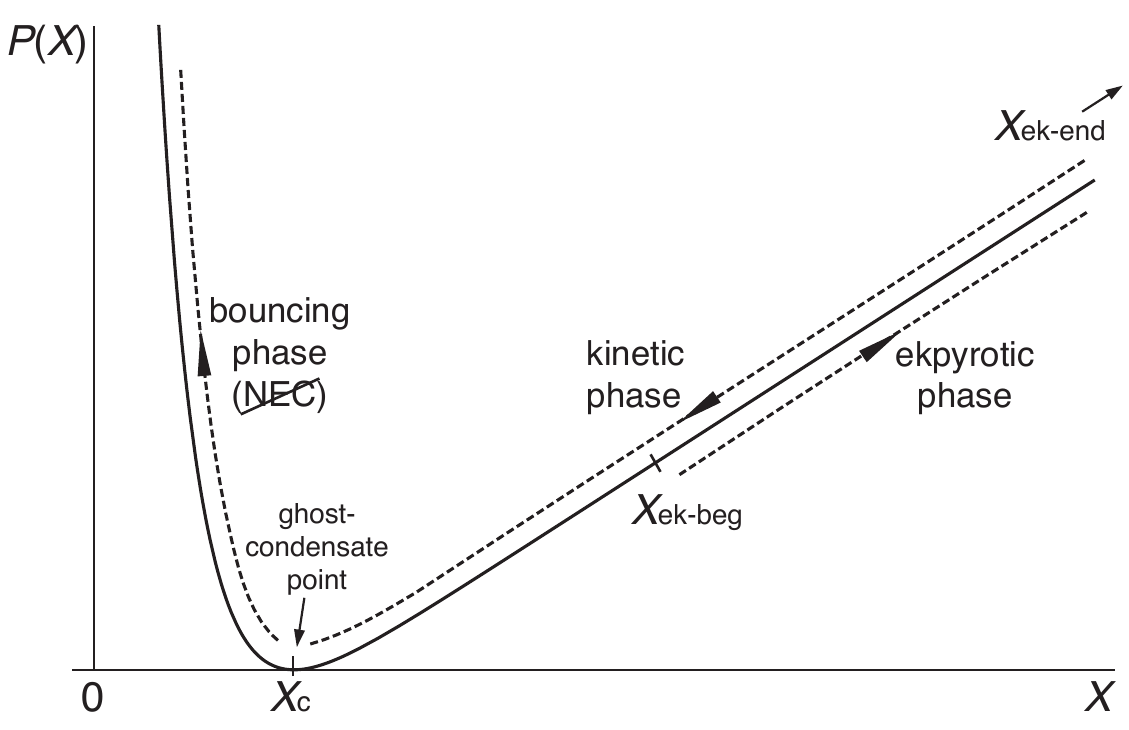}
\hspace{0.5cm}
\includegraphics[width=7cm]{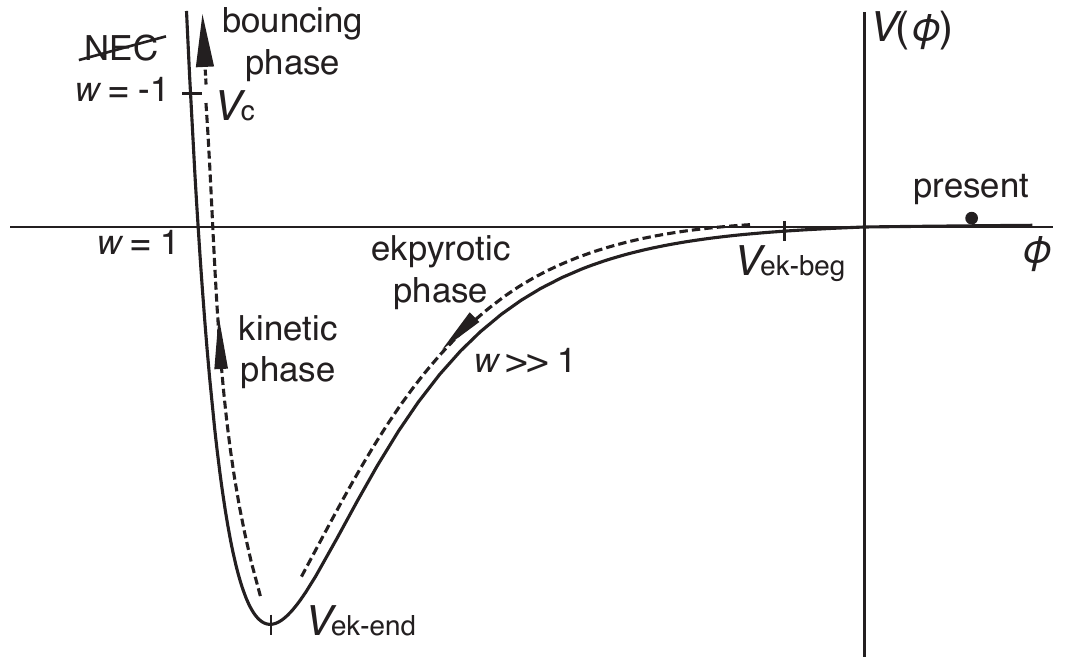}
\caption{\footnotesize Top: ghost condensate kinetic function $P(X)$. Bottom:
  ekpyrotic potential $V(\phi)$. Plots obtained
  from Ref. \cite{Xue:2011nw}.}
\label{fig:ekpyroticghost}
\end{center}
\end{figure}

If $X$ is a constant and $P_{,X}=0$ at $X=X_\mathrm{c}$, the equation
of motion yields the solution
\begin{equation}
\phi=\sqrt{2X_\mathrm{c}}\,t\,.
\end{equation}
Given that
\begin{equation}
\rho+P=2XP_{,X}\,,
\end{equation}
and the constraint $X>0$, a violation of the NEC can take place if
$P_{,X}<0$ in some interval of the values of $X$.  This, the condition
that $P_{,X}=0$ at $X=X_{\mathrm c}$ and requiring that $P_{,XX}>0$ at
$X=X_{\mathrm c}$ in order to prevent the existence of ghosts implies
that the function $P(X)$ should have a local minimum at $X_{\mrm c}$,
(see Fig.~\ref{fig:ekpyroticghost}).  This
construction was employed in the so-called $\Ka-$bounce \cite{AP07},
new Ekpyrotic \cite{Buchbinder:2007tw} and 
matter bounce \cite{Lin:2010pf} scenarios. Unfortunately, the ghost
condensate phase in models such as this cannot be smoothly connected
with a branch $P_{,X}>0$ at $X=0$.  Ghost condensate type models
therefore do not admit a stable Poincaré-invariant vacuum state and
are thus severely flawed. Consequences of this instability were
for instance demonstrated explicitly in the case of the new Ekpyrotic
scenario in Refs.~\cite{Kallosh:2007ad,Xue:2010ux}.

\paragraph*{$-$ {\it Ekpyrotic potential and ghost condensation}\\\\}

As was mentioned earlier, the contracting phase
of a bouncing cosmology is generically unstable under the growth of
anisotropies and leads to chaotic mixmaster oscillations unless a
period of ekpyrotic contraction with $w>1$ is invoked or if the
contraction is sufficiently brief. The smooth transition from
ekpyrotic contraction to expansion through a non-singular bouncing
phase relying on the NEC violation (with $w<1$) was first studied in
the new Ekpyrotic model \cite{Buchbinder:2007tw}.
It is obtained with a single scalar field rolling down a steep
negative potential during the ekpyrotic phase and then undergoing a
ghost condensation. In this approach the ghost condensate Lagrangian
is thus supplemented with a potential term $V(\phi)$.  The function
$P(X)$ realizing the ghost condensate phase, and the ekpyrotic
potential $V(\phi)$ are depicted in Fig.~\ref{fig:ekpyroticghost} As
shown for instance in~\cite{Xue:2011nw} however, this model suffers
from a gradient instability and from the regrowth of the initial
anisotropy during the bouncing phase. In addition, the absence of a
Lorentz-invariant vacuum remains, as in the ghost-condensate model.
It also predicts a blue spectrum of curvature perturbations.

\paragraph*{$\bullet$ {\bf Conformal Galileons}\\\\}

 Instead of realizing NEC-violations by relying on
Lagrangians that are restricted to general functions of the {\it
 kinetic term} only, it is possible to construct yet more general
NEC-violating theories with Lagrangians that exhibit couplings between
various sca\-lar field derivative terms \cite{Deffayet:2013lga}.
 Such scalar field theories are
called Galileon theories and have drawn much interest due to the fact
that they naturally admit self-accelerating solutions. In Galileon
theories, the scalar Lagrangian involves higher derivative
interactions with at most second order derivatives in the equations of
motion and is invariant under (possibly conformal or DBI-conformal)
Galilean transformations.  Galileon theories are a subclass of the
theory of Generalized Galileons which are described by the most general
scalar-tensor (Horndeski) action leading to second order equations of
motion. Denoting the scalar field by $\pi(x)$, it reads
\begin{align}
S_{\mrm H}=&\int \mrm{d}^4x \sqrt{-g}\bigg\{
P(\pi,X) + G_\Box (\pi,X)\Box \pi \cr
&\hskip-0.4cm + G_\mathcal{R}(\pi,X)\mcl R+G_{\mathcal{R}_{,X}}\lsb\lb \Box\pi\rb^2-\lb
\nabla_{\mu}\nabla_{\nu}\pi\rb^2\rsb \cr
& + G_G\lb\pi,X\rb G_{\mu\nu}\nabla^{\mu}\nabla^{\nu}\pi
-\displaystyle \frac{G_{G_{,X}}}{6} \times\nn\\
&\hskip-0.2cm \displaystyle \left[ \lb \Box
\pi\rb^3-3\Box\pi\lb\nabla_{\mu}\nabla_{\nu}\pi\rb^2+2\lb\nabla_{\mu}\nabla_{\nu}\pi\rb^3
\right] \bigg\},
\label{Hornd}
\end{align}
with $\mathcal{R}$ the scalar curvature, the functions
$G_\Box$, $G_\mathcal{R}$ and $G_G$ being arbitrary functions
of the field $\pi$ and its kinetic energy $X$; the last
two functions make manifest the non minimal couplings with the
gravitational sector.

Conformal formulations of Galileon theories are particularly
advantageous because the 4D conformal group reduces to the Anti de
Sitter, Minkowski and de Sitter symmetry groups for particular
solutions of the dilaton equation of
motion \cite{Creminelli:2012my}.
More importantly, whereas superluminal propagation of per\-tur\-bations is
common in higher derivative theories such as Galileons, in conformal
Galileon theories~\cite{Creminelli:2012my}, perturbations of the
scalar can be shown to travel with a speed at most equal to the speed
of light in the entire phase space as long as matter fields are
excluded~\cite{Easson:2013bda}.  In~\cite{Creminelli:2012my} a simple
and almost viable example is provided, with the theory described by
the Lagrangian
\begin{equation}
\mathcal
L=f^2\ex^{2\pi}(\partial\pi)^2+\frac{f^3}{\Lambda^3}(\partial\pi)^2
\Box\pi+\frac{f^3}{2\Lambda^3}(1+\alpha)(\partial\pi)^4
\label{galileonlagrangiandeformed},
\end{equation}
where $f$, $\Lambda$ and $\alpha$ are constant.
Provided $\mcl L$ features a negative
kinetic term, as is the case in Eq.~(\ref{galileonlagrangiandeformed}), this
theory admits a time-dependent de Sitter solution,
\begin{equation}
e^\pi=\frac{1}{-H_0t},\qquad \text{with} \quad
H_0=\frac{2}{3}\frac{1}{1+\alpha}\frac{\Lambda^3}{f},
\end{equation}
with $\Lambda$ the theory's strong coupling scale and violates the
NEC.  The sound speed is subluminal for $0<\alpha<3$ but the NEC
violating solution, as is the case for the ghost condensate cannot be
smoothly connected with a Lorentz-invariant vacuum solution.  This
theoretical setup nevertheless does bare a particularly interesting
aspect which was dubbed the
``Galileon Genesis'', namely that the cosmological solution displayed
here is an attractor solution. This allows the possibility to have
either emerging cosmological evolution or bouncing solutions.

\subsection{Semi-classical and quantum bounces}

Semi-classical models are those involving quantized scalar fields
in classical spacetimes.  The vacuum state being
ill defined on a curved back\-ground, except in the adiabatic limit of
slowly varying scale factor, a regularization or renormalization
scheme is required to cure such semi-classical theories of the
infinities that arise in the formal expression for the stress-energy
tensor.  These infinities are associated with the inability to
properly define the creation and annihilation operators and thereby
unambiguously remove the infinite vacuum term in the usual way.
Renormalization results in the inclusion of higher order curvature
(counter-) terms in the Einstein-Hilbert
action leading to new terms in the Friedmann
equations and to the possibility of constructing singularity avoiding
cosmologies.

A more ambitious approach to describe quantum gravitational
effects is string theory. The full action
of superstring theory possesses scale factor duality and time reversal
symmetry, which can be used to construct
non-singular cosmologies.  These require a {\it branch
  change} that smoothly interpolates between contracting and expanding
spacetimes.  This is the well-known {\it pre-big-bang}
cosmology\footnote{At low energy, the four dimensional effective
  action of the ekpyrotic model is
  equivalent to a modified version of the pre-big bang
  model.}~\cite{GV03}.
In its original version this model consists of only the dilaton field
and the metric.  At tree level, it can be shown that a contracting
cosmology in the string frame corresponds to an expanding
cosmology in the Einstein frame.  These two frames are simply related
by a conformal transformation.  It is therefore not clear whether to
identify such a tree-level cosmology with a {\it bouncing} cosmology
{\it per se}.  However, with the inclusion of loop
corrections, it is possible to show that the cosmology is indeed
non-singular,
and it is then plausible that one identifies this non-singular
evolution with {\it branch changing}.

Another possibility to smooth
out the curvature singularity is the inclusion of a coupling to a
matter or radiation fluid in the tree level effective 4D action of
string theory~\cite{FFPP03,Tsujikawa:2003pn}. In order to study the
propagation of cosmological perturbations through a bounce, it
can either be modeled by a discontinuity
across a spacelike hypersurface, in which case it is singular, and the
behavior of cosmological perturbations transfered through the bouncing
phase depends on how the Israel junction conditions are
implemented, or, alternatively, if one smooths the curvature
singularity by including
higher order corrections in the effective action, then it becomes
possible to actually follow perturbations through the
bounce.

Another string way to a non singular cosmology involves the motion
and interactions of
higher dimensional (mem)branes in yet higher dimensional bulk
geometries.  While initially, models were
based on branes embedded in, \eg 5-dimensional bulk
geometries, more recent models, based on either
heterotic M-theory or on the compactification of 10-dimensional type
II A/B superstring theory manifolds, and on the stabilization of the
``moduli'' fields\footnote{Moduli fields are fields that appear after
  compactification. They are identifiable with the (\apriori unfixed) sizes and
  shapes of cycles in the higher dimensional Calabi-Yau manifold which
  the theory lives on.}, have led to interesting brane dynamics in warped
parts of the geometry.  In general, these constructions lead to
additional terms in the Friedmann equations or unconventional kinetic
terms for the inflaton field that can lead to a period of inflation,
to bouncing branes and to cyclic cosmologies.  The {\it ekpyrotic}
model~\cite{KOST01} is one realization of brane
cosmology based on heterotic M-theory while examples of bouncing
cosmologies in warped string compactifications can be found in \eg
\cite{Kachru:2002kx,Easson:2007fz}.

A final option, also fully quantum, consists in assuming the energy
scale at which the bounce takes place to be sufficiently small that a
Wheeler de Witt treatment of quantum cosmology would be appropriate.
Although one naturally
faces measurement questions in such a context, there exists ways to treat
both background and perturbations on an equal footing; a
full set of predictions to compare with current data is however not yet
available since only toy models have been written down, but it would
seem that a consistent model should be attainable in the near future
(see Ref.~\cite{PPP07} and references therein).

\section{Conclusions}

Although inflation appears to largely dominate the field of
primordial cosmology, due, in particular, to the fact that many
implementations have predicted consequences rather similar to
the presently available observations, bouncing alternatives are not
entirely ruled out. In addition, a contracting phase
followed by a bounce can solve the primordial singularity problem,
which renders such models attractive and worth investigating.
It must be conceded however that, currently, most of bouncing models
have difficulties, either because they demand a complicated
theoretical framework or because their predictions disagree with
cosmological data.

From a purely theoretical standpoint, one might argue that bouncing cosmologies
relying on either ghost condensates or unknown quantum gravitational
effects \cite{Battefeld:2014uga} in order to successfully avoid the 
classical singularity should, in view of Occam's razor, be
disfavored when compared to much simpler inflationary models involving
perfectly well-behaved scalar fields \cite{Martin:2014vha}. One
should nevertheless remember that the singularity problem of
big-bang cosmology will need to be addressed at some stage,
and its solution, however contrived it may look from our perspective,
may end up being quite natural within a few decades. In
other words, one should not refute a theory or
a paradigm on philosophical grounds but instead on whether it is able
to answer as of yet unanswered physical questions and 
whether it agrees with the data or not.

From an observational perspective, given increasingly stringent
constraints imposed by present day cosmological data, many bouncing models are
under pressure as they naturally predict either exactly
scale-invariant scalar perturbations, or even slightly blue spectra.
Reddening the spectrum often demands
new components which may contribute in a non negligible way
and produce unobserved isocurvature mo-
des. Furthermore, the
bounce itself, involving either NEC violating fields or positive
spatial curvature, might induce large non Gaussianities
\cite{Gao:2014eaa}.  Present day bouncing models scarcely agree with
all existing observational constraints but it must be noted that neither do most
inflationary models~\cite{Martin:2014vha} . More work is needed to
reach definite conclusions, perhaps along the lines of purely quantum
models?

\section*{Acknowledgements}
PP wishes to thank Diana Battefeld with whom he co-authored
Ref.~\cite{Battefeld:2014uga} which has been inspirational to the present 
review.
\bibliography{references}

\end{document}